\documentclass[fleqn,usenatbib]{mnras}

\usepackage{newtxtext,newtxmath}

\usepackage[T1]{fontenc}
\usepackage{ae,aecompl}

\usepackage{graphicx}	
\usepackage{amsmath}	
\usepackage{amssymb}	

\title[K2-133]{Validation of a Temperate Fourth Planet in the K2-133 Multi-planet System}

\author[R. Wells, K. Poppenhaeger and C. A. Watson]{
R. Wells$^{1}$\thanks{E-mail: \href{mailto:rwells02@qub.ac.uk}{rwells02@qub.ac.uk}},
K. Poppenhaeger$^{1,2,3}$ and
C. A. Watson$^{1}$
\\
$^{1}$Astrophysics Research Centre, Queen's University Belfast, Belfast BT7 1NN, UK
\\
$^{2}$Leibniz-Institute for Astrophysics Potsdam (AIP), An der Sternwarte 16, D-14482 Potsdam, Germany\\
$^{3}$Institute for Physics and Astronomy, University of Potsdam, Karl-Liebknecht-Str. 24/25, D-14476 Potsdam, Germany
}
\date{Accepted XXX. Received YYY; submitted in original form ZZZ}

\pubyear{2018}

\begin{document}
\label{firstpage}
\pagerange{\pageref{firstpage}--\pageref{lastpage}}
\maketitle

\begin{abstract}
We present follow-up observations of the K2-133 multi-planet system. Previously, we announced that K2-133 contained three super-Earths orbiting an M1.5V host star -- with tentative evidence of a fourth outer-planet orbiting at the edge of the temperate zone. Here we report on the validation of the presence of the fourth planet, determining a radius of $1.73_{-0.13}^{+0.14}$ R$_{\earth}$. The four planets span the radius gap of the exoplanet population, meaning further follow-up would be worthwhile to obtain masses and test theories of the origin of the gap. In particular, the trend of increasing planetary radius with decreasing incident flux in the K2-133 system supports the claim that the gap is caused by photo-evaporation of exoplanet atmospheres. Finally, we note that K2-133~e orbits on the edge of the stars temperate zone, and that our radius measurement allows for the possibility that this is a rocky world. Additional mass measurements are required to confirm or refute this scenario.
\end{abstract}



\section{Introduction}
M-dwarf stars provide the best opportunity to study potentially habitable planets in the near-future. They are the most common stars in the Galaxy and have been estimated to host a rate of 0.45 small ($1-2 R_{\earth}$) temperate planets per star, using the recent Venus and early Mars boundary criteria \citep{2015ApJ...807...45D}. Due to their small stellar radii, transit signals of planets orbiting M-dwarfs are larger than for those around earlier-type stars. In addition, the orbital periods of planets in the temperate zones of M-dwarfs are very short (typically days to a few weeks), making their detection and characterisation more accessible. Similarly, the lower masses of these stars allow more accurate mass measurements of their planets. The greater transit signals also improve the studies of the atmospheres of these worlds via techniques such as transmission spectroscopy, potentially allowing searches for biosignatures that are currently extremely challenging for Earth-like planets in Earth-like orbits around solar-like stars \citep{2014Icar..242..172I}.

A multi-planet system was discovered around K2-133 -- an early M-star, using data from campaign 13 of the Kepler-K2 mission \citep{2018MNRAS.473L.131W}. The system consists of three confirmed super-Earths increasing in size with distance from the star -- possibly caused by photo-evaporation of their atmospheres. A candidate fourth super-Earth was also stated in the discovery paper, on a longer period orbit. In this work, through a careful re-analysis of the K2 data, we statistically validate this planet candidate with a period of 26.6 days and a radius of $1.73_{-0.13}^{+0.14}$ R$_{\earth}$, which orbits this quiet M-dwarf close to the optimistic habitable zone. The radius of this planet means it is likely to have a gaseous envelope \citep{2015ApJ...801...41R}, however we cannot rule out a rocky composition with the current uncertainty. Future photometric observations with higher cadence are needed to further constrain the radius.

Since the discovery, new data has been acquired allowing a more detailed characterisation of the system. In Section~\ref{methods} we present AO imaging, NIR spectra and Gaia DR2 data which further constrain the properties of the host star and therefore also the planets. In Section~\ref{validation}, we give the results from a statistical test of the probability that the candidate planet is real, and provide its properties. The planet is discussed in detail in Section~\ref{sec:discussion}, considering its potential for habitability and scope for future observations. The planetary system as a whole is also discussed in the context of the known exoplanet radius gap.

\section{Methods}\label{methods}

\subsection{Keck AO imaging}
High resolution imaging can be used to rule out background eclipsing binary scenarios which can mimic transit signals, and also to identify blended sources which cause us to underestimate transit depths. K2-133 was observed by PI Crossfield on the night of August 20 2017, using the NIRC2 adaptive optics imager on the 10m Keck-II telescope. The data were obtained using the narrow Br-$\gamma$ filter ($\lambda_\mathrm{eff} = 2.157 \mu m$, $\Delta \lambda = 0.043 \mu m$), combining 9 dithered images with integration times of 10.3 seconds each. More details of the setup used can be found in \citet{2018AJ....155...10C}, \citet{2016ApJS..226....7C} and \citet{2018AJ....156..277L}. The data are publicly available at the ExoFOP website\footnote{\url{https://exofop.ipac.caltech.edu/k2/edit_target.php?id=247887989}}. 

\begin{figure}
\begin{center} 
\includegraphics[width=.47\textwidth]{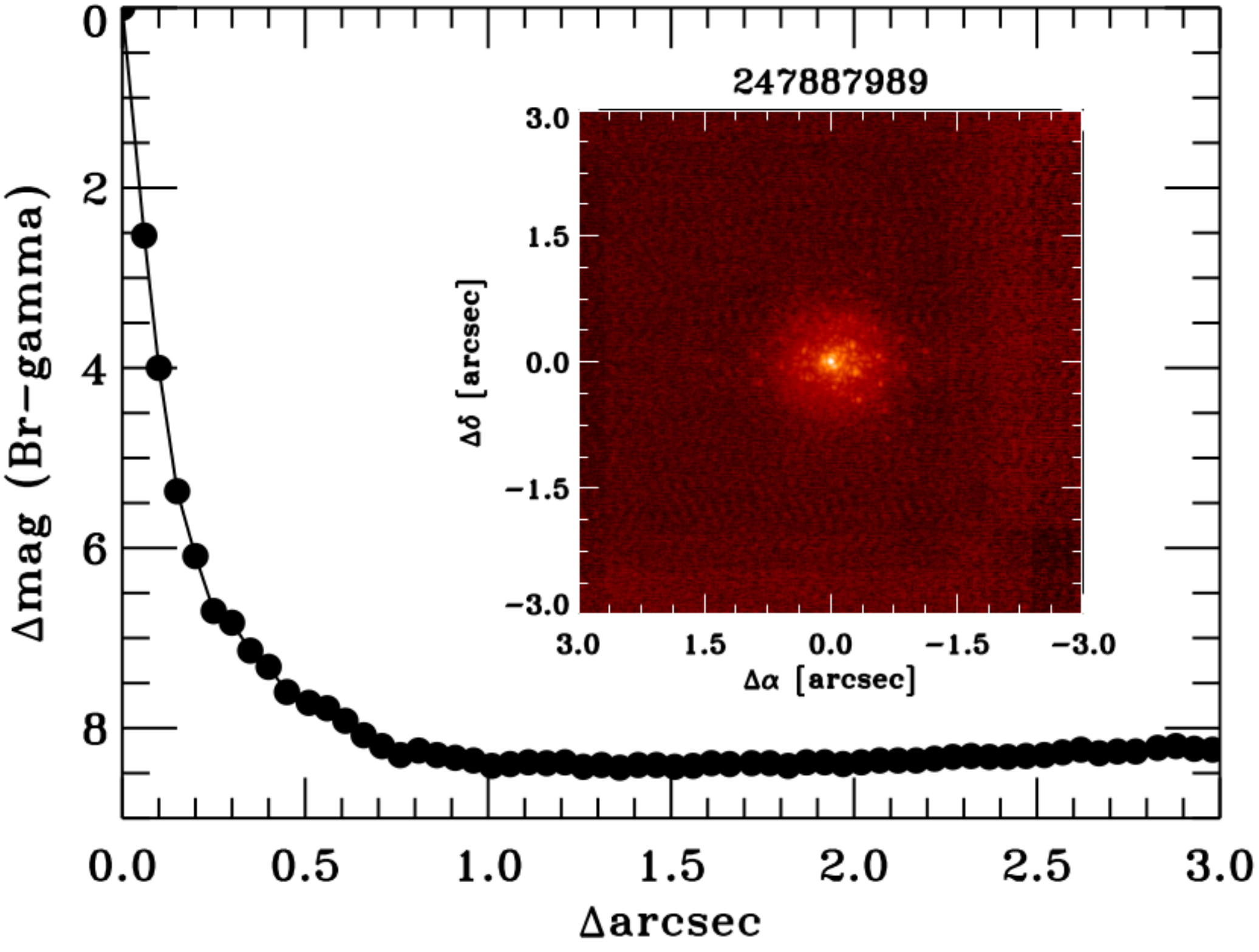} 
\caption{Contrast sensitivity curve for K2-133, observed with Keck/NIRC2 in the Br-$\gamma$ filter. The data points and curve represent the $5 \sigma$ contrast limits in $\Delta$mag as a function of distance from K2-133. The reduced, co-added image is inset, where no secondary source is detected. This figure is reproduced with permission from PI Crossfield. \label{fig:AO}} 
\end{center} 
\end{figure}

Fig.~\ref{fig:AO} shows the final image after flat-fielding, sky subtraction and then co-adding the dithered images. The $5 \sigma$ sensitivity curve is shown, resulting from injection and recovery tests of fake sources at different distances from the main source. K2-133 is clearly seen in centre and no extra sources are detected above the contrast curve, i.e. to a contrast of 7.7 mag at $0.5^{\prime\prime}$ and 8.4 mag further than $1^{\prime\prime}$.

\subsection{Spectroscopic follow-up}
We obtained zJ and HK NIR spectra of K2-133 using LIRIS \citep{1998SPIE.3354..448M} at the 4.2m William Herschel Telescope. We also obtained spectra of HD\,27267, a nearby rapidly-rotating A0V star for use as a telluric standard. The observations were taken in service mode on October 09 2017 (proposal SW2017b04). K2-133 and HD\,27267 are located $5.4^\circ$ apart on the sky, resulting in a ca.~0.03 difference in airmass between the targets at the time of observation. We used the ABBA nod pointing pattern with exposures of 25 and 12 seconds for K2-133 for the zJ and HK grisms, respectively. Four exposures were taken for each grism and pointing position. The spectra were reduced and combined using the {\sc lirisdr} {\sc iraf} package\footnote{{\url{https://github.com/jaacostap/lirisdr}}}, and wavelength calibrated using spectra of Argon arc lamps. The K2-133 spectra were then corrected for tellurics using the extracted A0V spectra and the {\sc idl} routines of \citet{2003PASP..115..389V}.

\begin{figure*}
\begin{center} 
\includegraphics[width=.95\textwidth]{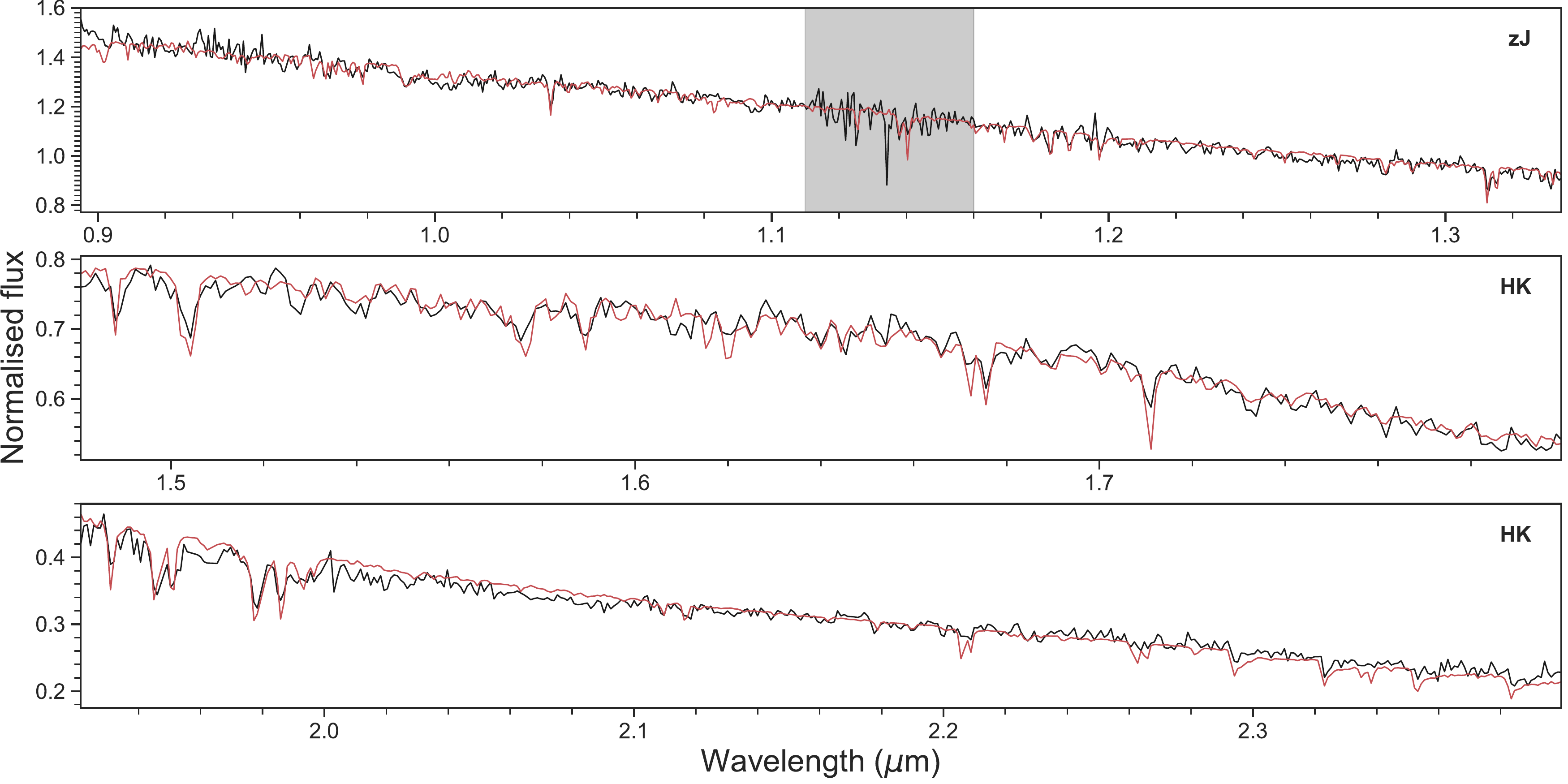}
\caption{LIRIS NIR spectrum of K2-133 (black) with best-fitting template of HD\,42581 over-plotted (red). The top panel shows the zJ spectrum and the other two display the HK grism, separated by a telluric feature between 1.8 and 1.9 $\mu m$. The grey region between 1.11 and 1.16 $\mu m$ is a leftover telluric region that was masked in our analysis. \label{fig:spectrum}} 
\end{center} 
\end{figure*}

While bulk tellurics were corrected, some features were left in the resulting spectrum, possibly due to slightly differing airmass between the target and comparison star or atmospheric changes between exposures. We therefore masked regions of leftover telluric features in the corrected spectrum of K2-133 ($1.11-1.16$, $1.34-1.48$ and $1.80-1.92$ $\mu m$) to not interfere with our interpretations. The corrected spectrum can be seen in Fig.~\ref{fig:spectrum}, split into three parts bounded by two of the masked telluric regions. 

\begin{figure}
\begin{center} 
\includegraphics[width=.45\textwidth]{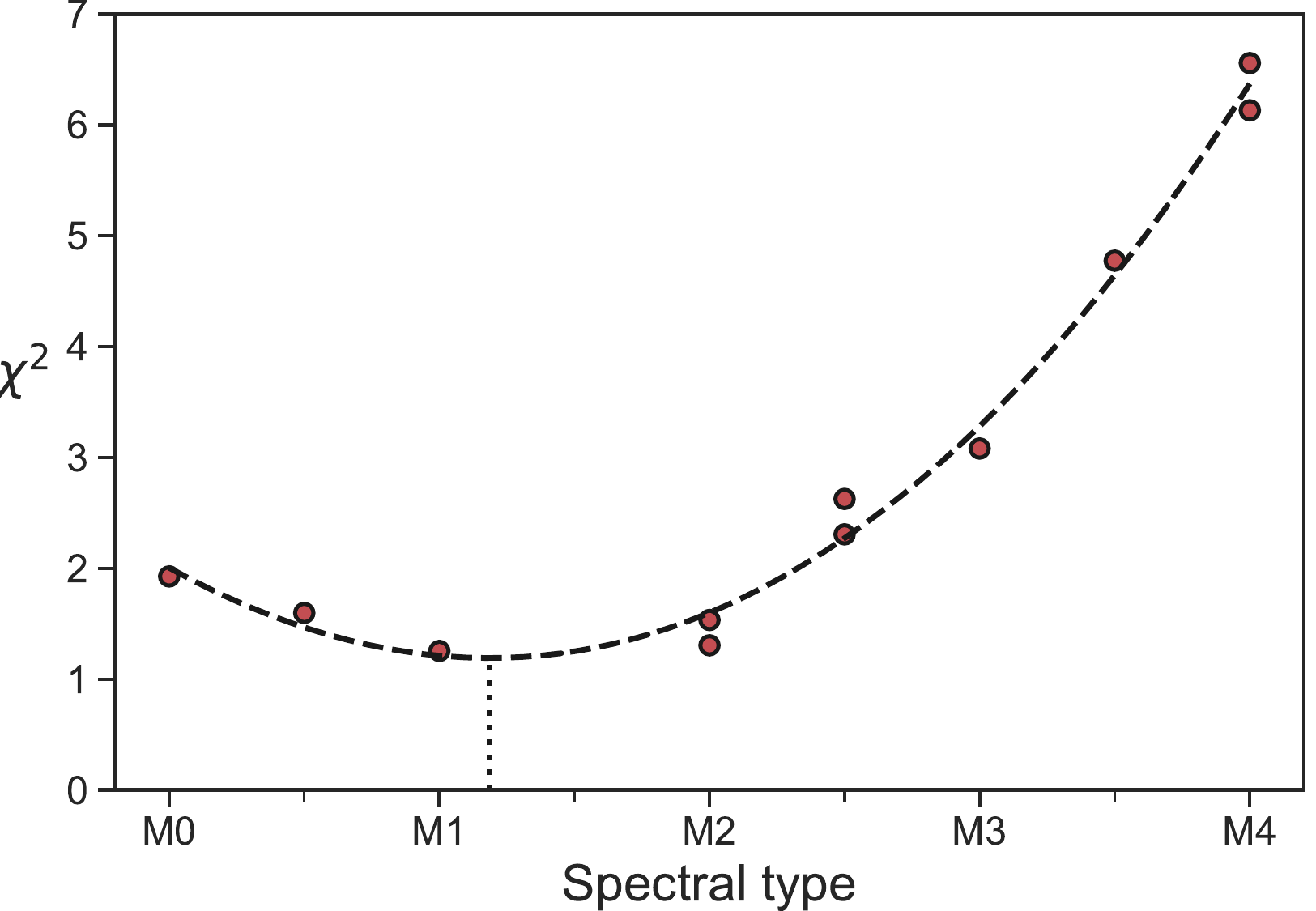}
\caption{Results of the $\chi^2$ test for each M-dwarf template spectrum compared to the K2-133 data. A second order polynomial was fitted to the $\chi^2$ values, with a minimum between spectral types M1 and M2. Note that two templates are available in the IRTF library for spectral types M2, M2.5 and M4. \label{fig:chisq}} 
\end{center} 
\end{figure}

The corrected K2-133 spectrum was compared to template M-dwarf spectra from the IRTF Spectral Library\footnote{\url{http://irtfweb.ifa.hawaii.edu/~spex/IRTF_Spectral_Library/index.html}} \citep{2005ApJ...623.1115C,2009ApJS..185..289R}. The template spectra were first binned to the same wavelength using the \texttt{pysynphot} Python package \citep{2013ascl.soft03023S}. Then each template was shifted and scaled to best match the K2-133 spectrum, and finally a $\chi^2$ test was conducted between them. The result of the $\chi^2$ tests can be seen in Fig.~\ref{fig:chisq}, showing increased agreement of the data with templates of spectral types M2/M1. This is close to the M1V spectral type found in \citet{2018MNRAS.473L.131W} and is consistent with the closer distance from Gaia (see Section~\ref{sec:gaia}). The IRTF template spectrum with the lowest $\chi^2$ was HD\,42581 -- an M1V star which is over-plotted on Fig.~\ref{fig:spectrum}. An almost equally good $\chi^2$ was given by the M2V HD\,95735 template.

\subsection{Gaia DR2 \label{sec:gaia}}
Properties of the host star are further constrained by a precise parallax measurement from Gaia DR2 \citep{2018A&A...616A...1G}, providing a distance of $75.2 \pm 0.2$ pc \citep{2018AJ....156...58B}. This is consistent with the distance of $78 \pm 7$ pc reported in \citet{2018MNRAS.473L.131W}, but far more precise. The slightly closer distance to K2-133 supports the analysis of the NIR spectrum, suggesting the star is closer to an M2 spectral type than previously reported. 

We used the new distance to determine revised values of the stellar radius, mass and luminosity. We first calculated the absolute K-band magnitude using the apparent 2MASS K magnitude of $10.279 \pm 0.018$ \citep{2006AJ....131.1163S} and the Gaia-derived distance. We then found the bolometric correction, BC$_K$, using Equation~10 from \citet{2015ApJ...804...64M} (note the erratum \citealt{2016ApJ...819...87M}) with the apparent J and SDSS r magnitudes. We derived the bolometric luminosity directly from the absolute K mag and BC$_K$, assuming Solar metalicity and no K-band extinction. The effective temperature was computed using the relations of \citet{2015ApJ...804...64M}, and the stellar mass from \citet{2018arXiv181106938M}. The stellar radius was computed from the derived bolometric luminosity and effective temperature. The updated stellar values are given in Table~\ref{tab:stellar}, with uncertainties given at $1 \sigma$ confidence level from an MCMC analysis.

We verified these stellar properties by comparing them to stellar models. We utilised the \texttt{isochrones} Python package \citep{2015ascl.soft03010M}, using the MIST stellar isochrones \citep{2016ApJ...823..102C,2016ApJS..222....8D}, and the 2MASS JHK magnitudes and Gaia parallax as inputs. This resulted in stellar parameters of T$_{\mathrm{eff}} = 3770_{-75}^{+104}$ K, R$_{\star} = 0.440_{-0.013}^{+0.010}$ R$_{\sun}$, M$_{\star} = 0.461 \pm 0.014$ M$_{\sun}$ and L$_{\star} = 0.0351_{-0.0015}^{+0.0019}$ L$_{\sun}$. These are in good agreement with the values from the empirical relations, in particular, both the mass and radius agree better than $1 \sigma$.

\begin{table}
\caption{Refined stellar properties of K2-133 from Gaia astrometry, and their $1 \sigma$ uncertainties.}
\begin{tabular}{lrl}
\hline \hline
Property & Value & Unit \\
\hline
Gaia astrometry: & & \\
Ref. Epoch & 2015.5 & Jyr \\
R.A. & $70.1494$ & deg. \\
Dec & $25.0098$ & deg. \\
$\mu_{\mathrm{RA}}$ & $185.66 \pm 0.08$ & mas\,yr$^{-1}$ \\
$\mu_{\mathrm{Dec}}$ & $-46.36 \pm 0.04$ & mas\,yr$^{-1}$ \\
Radial velocity & $96.9 \pm 1.7$ & km\,s$^{-1}$ \\
Parallax & $13.27 \pm 0.04$ & mas \\
& & \\
Derived properties: & & \\
$T_{\mathrm{eff}}$ & $3655\pm 80$ & K \\
Distance & $75.2 \pm 0.2$ & pc \\
Stellar radius & $0.455 \pm 0.022$ & R$_\odot$ \\
Stellar mass & $0.461 \pm 0.011$ & M$_\odot$ \\
Stellar luminosity & $0.0332 \pm 0.0013$ & L$_\odot$ \\
Space velocity & $107.2 \pm 1.4$ & km\,s$^{-1}$ \\
\hline
\end{tabular}
\label{tab:stellar}
\end{table}

\subsection{Transit fitting} \label{sec:transit-fitting}
We have computed refined properties of the transiting planets, using the new stellar properties with the {\sc batman} transit modelling code \citep{2015PASP..127.1161K,2002ApJ...580L.171M} and \texttt{emcee} \citep{2013PASP..125..306F,2010CAMCS...5...65G}. We also fit for the limb-darkening coefficients using the three-parameter law and the \citet{2016MNRAS.455.1680K} sampling method. We note that the limb-darkening coefficients are not well defined, but they are important when estimating uncertainty in the inclinations of the planets. Since the original discovery, the K2 team has begun an effort to reduce each campaign with a uniform version of the reduction pipeline. We therefore use the reprocessed lightcurve in our analysis, with systematics removed using a modified version of the \texttt{Lightkurve} \citep{lightkurve} implementation of the {\sc k2sff} method \citep{2014PASP..126..948V}.

\begin{table*}
\caption{Refined planetary parameters of the system with uncertainties given at the $1 \sigma$ confidence level.}
\begin{tabular}{lccc}
\hline\hline
Property & Planet b & Planet c & Planet d \\
\hline
Fitted properties: & & & \\ [0.1cm]
$T_{0}$ (BJD$-2454833$) & $2988.3150_{-0.0015}^{+0.0016}$ & $2990.7686 \pm 0.0011$ & $2993.1709 \pm 0.0012$ \\ [0.1cm]
Period (days) & $3.07133 \pm 0.00011$ & $4.86784 \pm 0.00012$ & $11.02454_{-0.00035}^{+0.00036}$ \\ [0.1cm]
$R_{p}/R_{s}$ & $0.0270 \pm 0.0008$ &  $0.0323 \pm 0.0009$ & $0.0404_{-0.0010}^{+0.0008}$ \\ [0.1cm]
Inclination (deg.) & $87.60_{-0.15}^{+0.18}$ & $88.21_{-0.08}^{+0.11}$ & $89.40_{-0.08}^{+0.10}$ \\ [0.1cm]
& & & \\
Derived properties: & & & \\ [0.1cm]
Radius ($R_{\earth}$) & $1.340_{-0.076}^{+0.077}$ & $1.603_{-0.088}^{+0.090}$ & $2.003_{-0.107}^{+0.107}$ \\ [0.1cm]
Semi-major axis (au) & $0.03194_{-0.00026}^{+0.00025}$ & $0.04341_{-0.00035}^{+0.00034}$ & $0.07487_{-0.00060}^{+0.00059}$ \\ [0.1cm]
Impact parameter & $0.630_{-0.054}^{+0.052}$ & $0.640_{-0.047}^{+0.046}$ & $0.370_{-0.066}^{+0.055}$  \\ [0.1cm]
& & & \\
Non-linear LDCs & c$_2 = 3.4_{-1.9}^{+2.1}$ & c$_3 = -5.6_{-4.4}^{+3.5}$ & c$_4 = 2.9_{-1.9}^{+2.4}$ \\ [0.1cm]
\hline
\end{tabular}
\label{tab:planetparams}
\end{table*}

\section{Validation of planet e}\label{validation}
With our refined stellar properties and lightcurve, we revisited the planetary candidate transit signal of K2-133~e. We note the transit is now detected at a ${\sim} 13 \sigma$ significance, leading us to evaluate the probability that it is caused by a fourth planet in the system and is not due to random or correlated noise. In order to validate the candidate, we require a false-positive probability (FPP) of less than one per cent, which is typically used to validate transit events. 

We first note that K2-133 was located on Module 15, Output 2 (Channel 50) during observations. This channel has no reported detector irregularities from the K2 engineering test \citep{2016ksci.rept....1V}. We also inspected the background counts (pixels outside the aperture) for K2-133 and two nearby stars -- EPIC~247894072 and EPIC~247876252. We found no disagreements between background levels of the three stars and no signs of variability or trends, such as rolling band effects, which could affect the detected transit events. 

\subsection{Detrending}\label{sec:detrending}
\begin{figure}
\begin{center} 
\includegraphics[width=.45\textwidth]{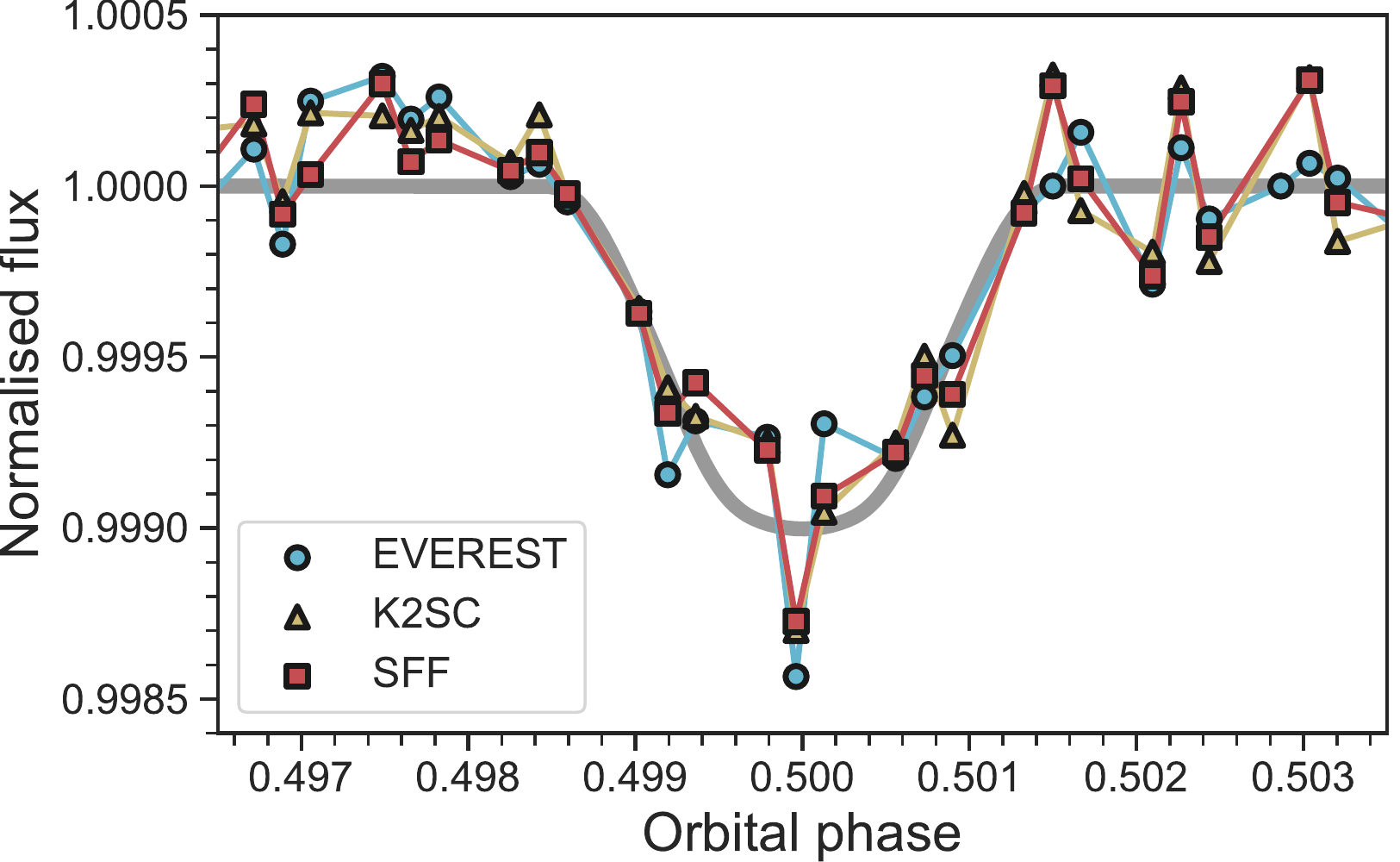}
\caption{Phase-folded lightcurve for the planetary candidate. Three different detrending methods are shown, as described in the main text. \label{fig:detphase}} 
\end{center} 
\end{figure}

Our first test of the planet candidate was to use multiple detrending methods and see whether the transit signal was apparent in each detrended lightcurve. We utilised the {\sc k2sc} algorithm \citep{Aigrain2016}, the {\sc everest} high-level science product \citep{2018AJ....156...99L} and our SFF lightcurve from Section~\ref{sec:transit-fitting}. We note that the {\sc everest} lightcurve is from detrending the original (non-reprocessed) K2 data. We find the same transit feature in all three lightcurves with good agreement between each method as seen in Fig.~\ref{fig:detphase}, indicating the transit signal is not due to a detrending artefact. For the rest of our analysis we use the SFF lightcurve because it has the best 6.5-hour combined differential photometric precision (CDPP) of 145 ppm and the least visible correlated noise. We further removed data points corresponding to an asteroid passing through the aperture at time (BJD$-2454833$) ca.~3018 days and any exposures with bad Kepler quality flags.

\subsection{Injection test}\label{sec:injections}
Our next goal was to test the sensitivity of our detection pipeline. To do this, we injected planetary transits into the full, non-detrended lightcurve with all four planets removed. The injected transits covered the parameter space of: 1-40 day orbital periods and transit depths of 400-1400 ppm. For reference, the candidate has an orbital period of 26.6 days and a depth of 1000 ppm. We used the {\sc batman} transit model \citep{2015PASP..127.1161K} with a super-sampling factor of 15, varying the parameters of orbital phase, orbital period, planetary radius and orbital inclination. The semi-major axis was computed using Kepler's third law. The eccentricity was fixed to zero, as this parameter does not largely effect K2 lightcurves due to the time sampling of 30 minutes. The stellar radius, mass and quadratic limb darkening coefficients ($\mu_1=0.5079, \mu_2=0.2239$ from \citealt{2010A&A...510A..21S}) were also kept fixed for each injected model. Each injected planet had a minimum of three transits present in the lightcurve.

\begin{figure*}
\begin{center} 
\includegraphics[width=.95\textwidth]{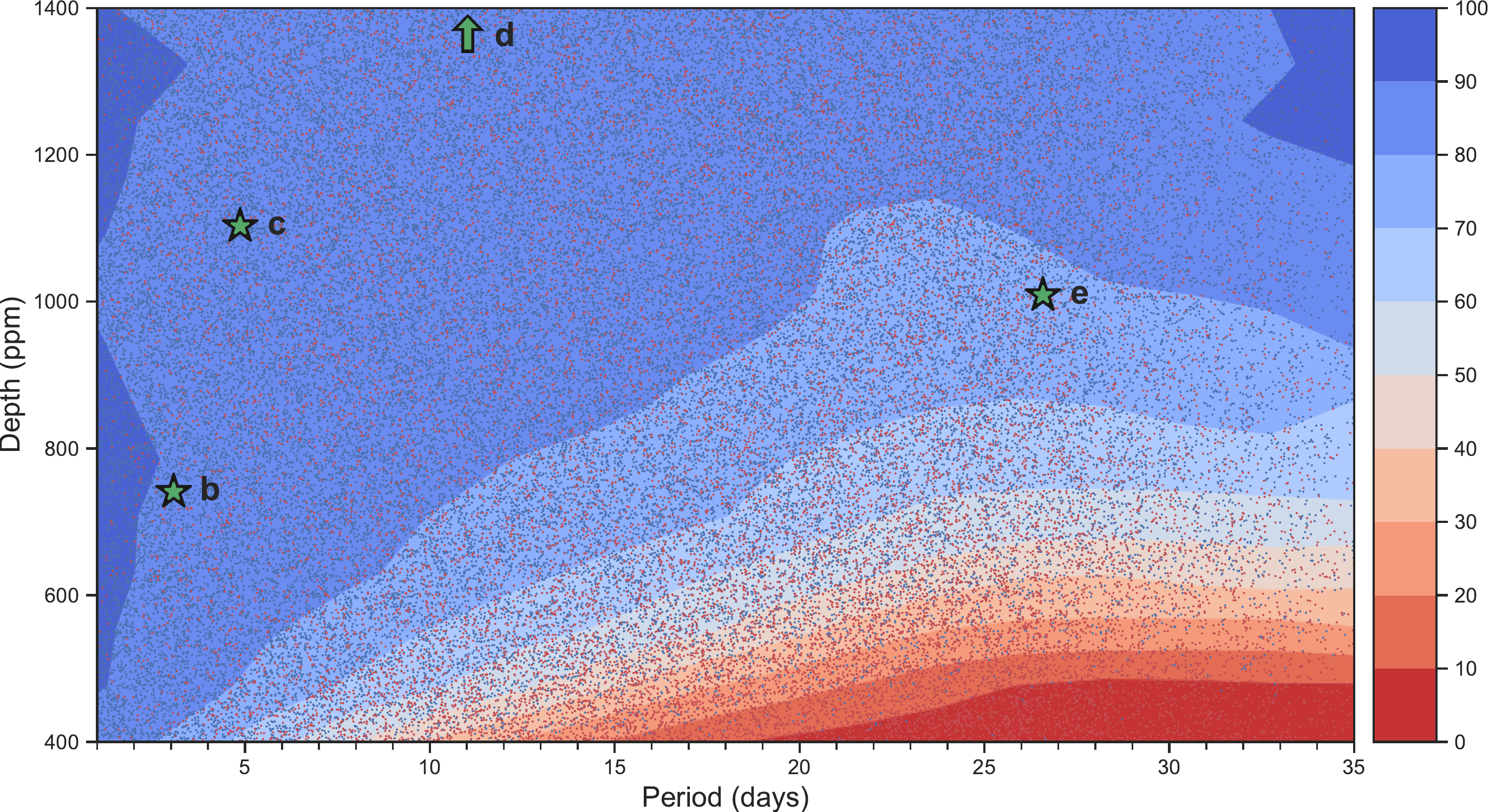}
\caption{Recovery rate from our injection test as a function of orbital period and transit depth. Individual recovered and missed injected planets are shown as blue and red points, respectively. The recovered percentage is over-plotted as smoothed contours, coloured from red (0\%) to blue (100\%). Planets b, c and e are marked with stars, while planet d is positioned off-grid at ca.~2000 ppm deep. \label{fig:injection}} 
\end{center}
\end{figure*}

To detect planets, we utilised the Box-fitting Least Squares (BLS) algorithm of \citep{2002A&A...391..369K} which searches for periodic box-shaped dips in a light curve. A number of trial periods are tested and each return a signal detection efficiency (SDE) signal-to-noise statistic. To recover an injected planet, we required the BLS SDE to be above 8 and the orbital period to be identified within 1\% of the injected value. This is very similar to the tests performed in \citet{2017AJ....154..224R}. We ran 64,000 injection tests which are summarised in Fig.~\ref{fig:injection}, showing the recovered percentage as a function of the injected transit depth and orbital period. There is a clear trend of increased recovery rate for shorter periods and larger transit depths, which greatly impact the BLS algorithm. Planets b, c and e are marked on the figure and reside in regions of 70-90\%. Planet d has a transit depth of 2000 ppm and orbital period of 11 days, placing it well into the 90+\% region. We therefore conclude that the K2 data is of high enough quality that we should be able to detect transit signals of planet~e.

\subsection{Bootstrap tests}
After deciding our detection pipeline was sufficiently sensitive to the properties of the planetary candidate, we next tested the probability that the transit signal could be caused by random or correlated noise. To do so, we first computed the SDE of the candidate transit signal in our refined lightcurve. We used the \texttt{PyBLS}\footnote{\url{https://github.com/hpparvi/PyBLS}} implementation of the BLS algorithm using a period range of $1-40$ days, a transit duration to period ratio range of $0.001-0.2$, 50,000 frequency bins and no binning of the phase folded lightcurves. We then subtracted the running median from the resulting power spectrum using a window size of 1500 points, resulting in a peak SDE value of $10.7$. We then removed the transit signal by subtracting our best-fit transit model, resulting in a lightcurve composed of a combination of red and white noise. The highest BLS SDE of this lightcurve was $4.8$, far below any detection limit. This lightcurve was used for all false positive tests.

\begin{figure}
\begin{center} 
\includegraphics[width=.45\textwidth]{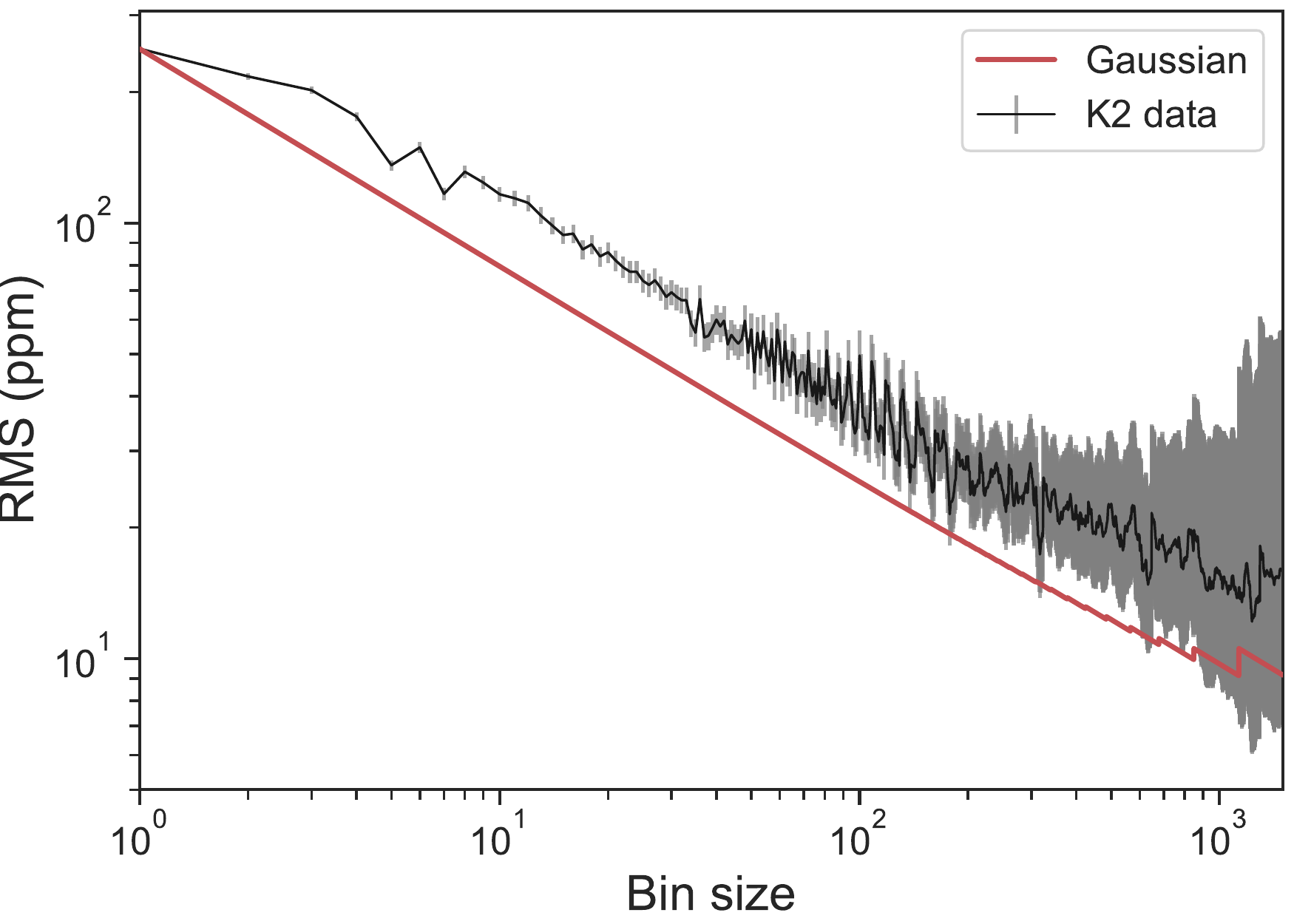}
\caption{RMS vs. bin size for our K2 residuals (black), and expected RMS if only white noise was present (red). \label{fig:red-noise}} 
\end{center} 
\end{figure}

A transit signal can be mimicked by both white and red noise, so we first checked how the level of correlated noise compared to the Gaussian white noise. We used the {\sc mc3} Python package \citep{2017AJ....153....3C} to compute the RMS of our data and the RMS expected from only white noise for a range of bin sizes. The result is plotted in Fig.~\ref{fig:red-noise}, where the red noise is roughly 40\% above the white noise. I.e. our lightcurve is white noise dominated but contains a non-negligible amount of red noise as well. Consequently, we decided to test the candidate transit signal against red and white noise separately.

\begin{figure}
\begin{center} 
\includegraphics[width=.45\textwidth]{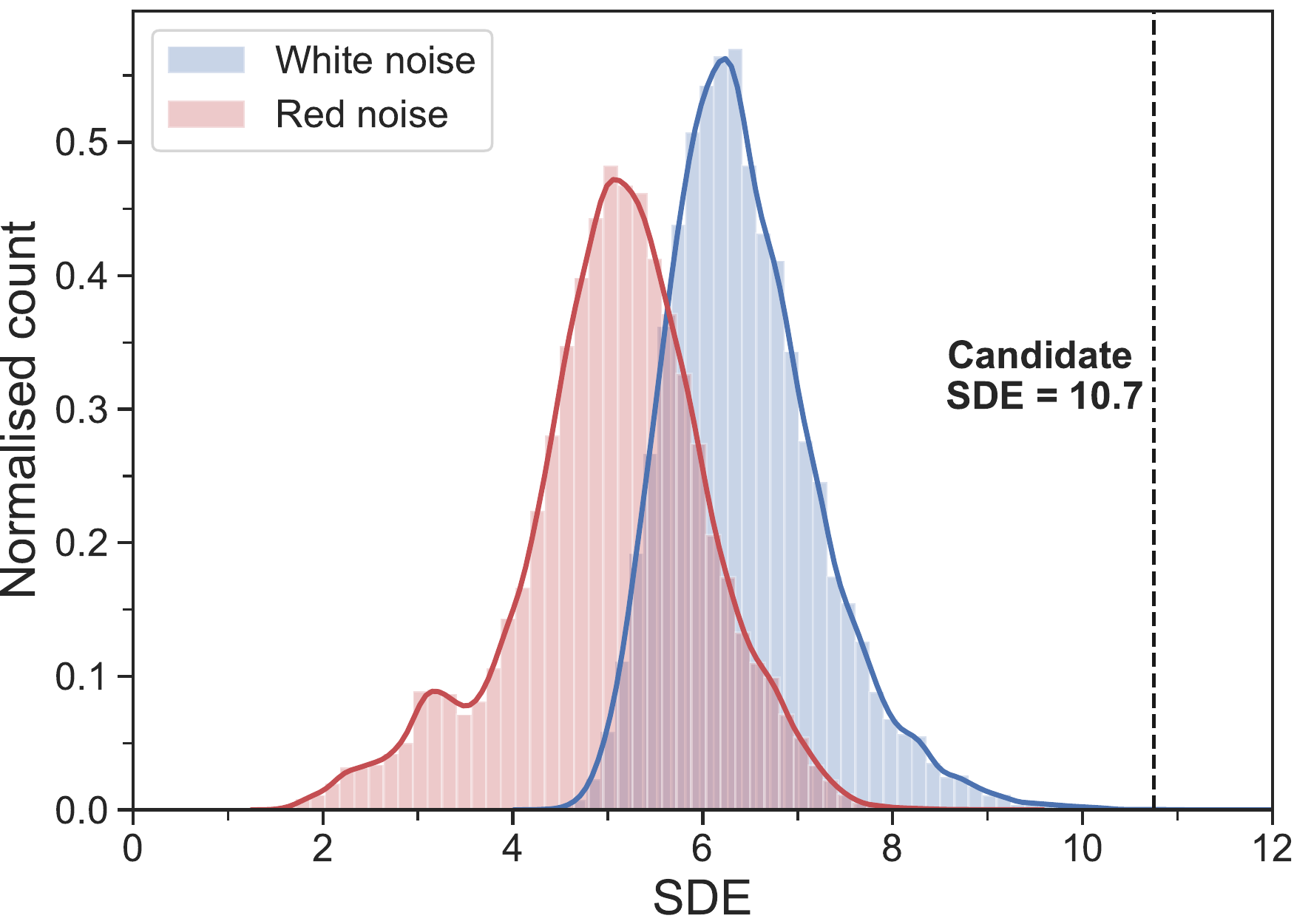}
\caption{BLS SDE distributions resulting from the bootstrap tests. The white noise and red noise distributions are shown in blue and red respectively. The SDE of the validated planet candidate is marked at 10.7. False-positive probabilities for both of these distributions are given in the text. \label{fig:bootstrap}} 
\end{center} 
\end{figure}

To test against white noise, we implemented a bootstrap test. For each test in the bootstrap, we randomly sampled the flux value at each cadence from the whole light curve, while keeping the times and flux errors the same. We repeated this 21,000 times, for each re-sample searching for periodic signals using the BLS algorithm with the same parameters as for the candidate. The resulting BLS SDE distribution is shown in Fig.~\ref{fig:bootstrap}, over-plotted with a kernel density estimation (KDE). We calculated the cumulative probability of the KDE at the SDE value of the candidate (10.7) and then the FPP was found by subtracting this value from unity, giving a FPP of 0.017\%. 

To test against correlated noise, we repeated this process with a different sampling method, this time using 50,000 re-samples. We opted for the bootstrapping method of \citet{1991Sci...253..390E}, used in the analysis of Roche tomograms by \citet{2001MNRAS.326...67W}. In short, we randomly selected data points from the entire light curve, placing them at their original positions in the new light curve. The flux uncertainty values were then reduced by the square-root of the number of times the data point was selected. I.e. if a data point was selected $n$ times the associated error value was reduced by $\sqrt{n}$; if a point was not selected at all then it was removed from the lightcurve. This method should show up correlated noise as periodic signals, given enough samples. The resulting distribution is again shown in Fig.~\ref{fig:bootstrap}, over-plotted with a kernel density estimation (KDE). The FPP was found to be $2.7 \times 10^{-6}\%$. 

\subsection{Astrophysical false positives}
We also tested the transit signal against eclipsing binary scenarios, using the {\sc VESPA} Python package \citep{2015ascl.soft03011M,2012ApJ...761....6M}. {\sc VESPA} computes likelihoods of the signal being caused by eclipsing binaries (EBs), hierarchical eclipsing binaries (HEBs) and background eclipsing binaries (BEBs), as well as planetary transits. Scenarios are also considered at twice the candidate orbital period. {\sc VESPA} then excludes systems that are inconsistent with the input data.

The SDSS, 2MASS and Kepler magnitudes alongside the Gaia parallax were used as inputs for the stellar populations. The fitted orbital period and planet-to-star radius ratio, transit light curve, Keck contrast curve and stellar RA and Dec were used as constraints to compute the likelihood of each scenario. We used a simulation sample size of 100,000 for each population and ran 10 bootstrap resamples. We obtained mean probabilities of 0.08, 2.28, 0.00 and 97.65 per cent for the EB, HEB, BEB and planetary scenarios, respectively. This equates to a FPP of 2.35 per cent. \citet{2012ApJ...750..112L} have shown that false-positives are less-likely in multi-planet systems, and the false-positive probability of a transit signal in a $2+$ planetary system is decreased by a factor of approximately 50. Applying this to our candidate reduces the FPP to ca.~0.05\%.

With both noise and astrophysical false-positive probabilities well below one per cent, we confirm the planet candidate, now named K2-133\,e.

\subsection{Properties}
\begin{figure*}
\begin{center} 
\includegraphics[width=.95\textwidth]{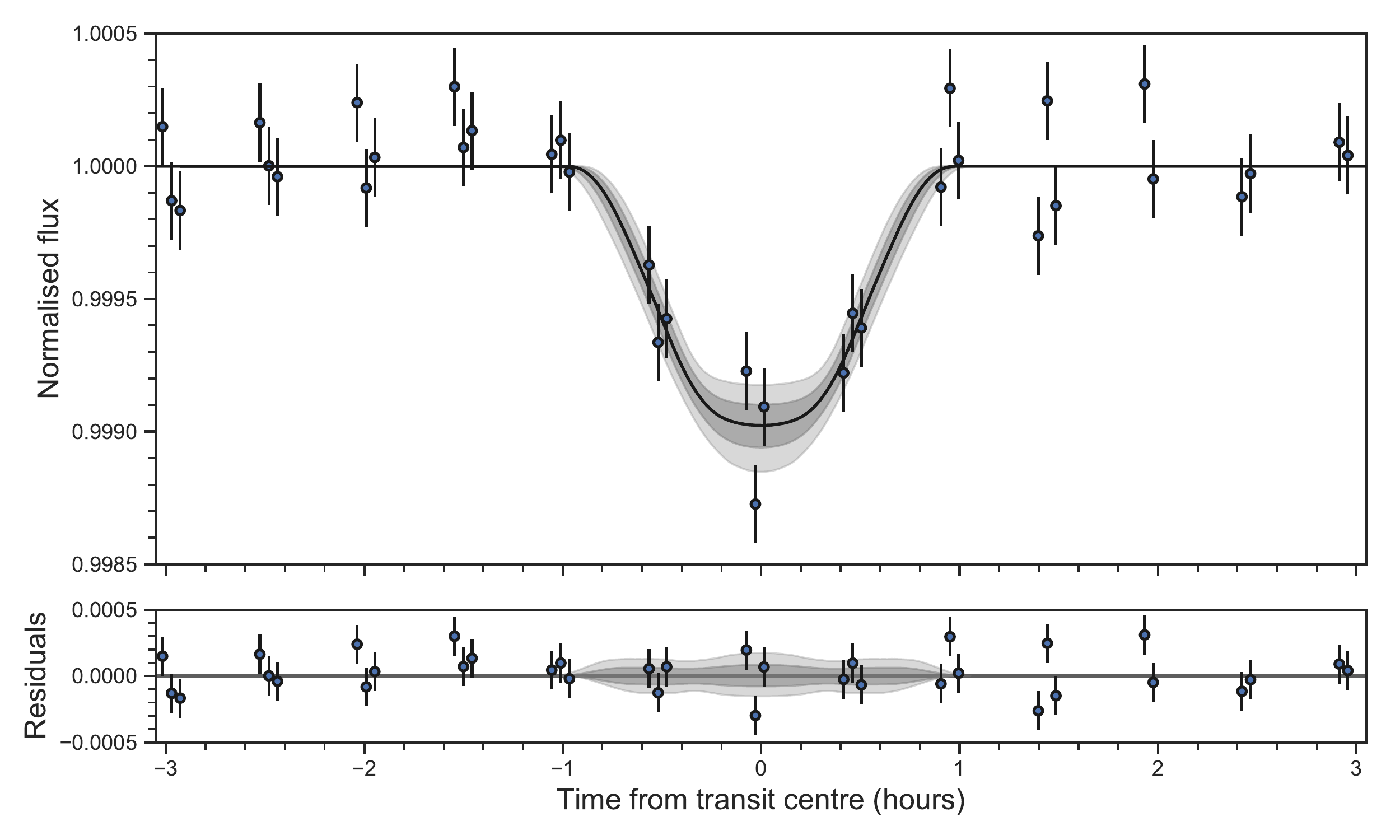}
\caption{Phase-folded light curve of planet~e resulting from the MCMC fit. K2 long-cadence data points are shown with errorbars from the 6.5-hour CDPP of the light curve. The model 1-- and 2--sigma uncertainties are shaded in grey around the median model plotted in black. Residuals to the transit model are plotted below. \label{fig:mcmcfit}} 
\end{center} 
\end{figure*}

\begin{figure}
\begin{center} 
\includegraphics[width=.47\textwidth]{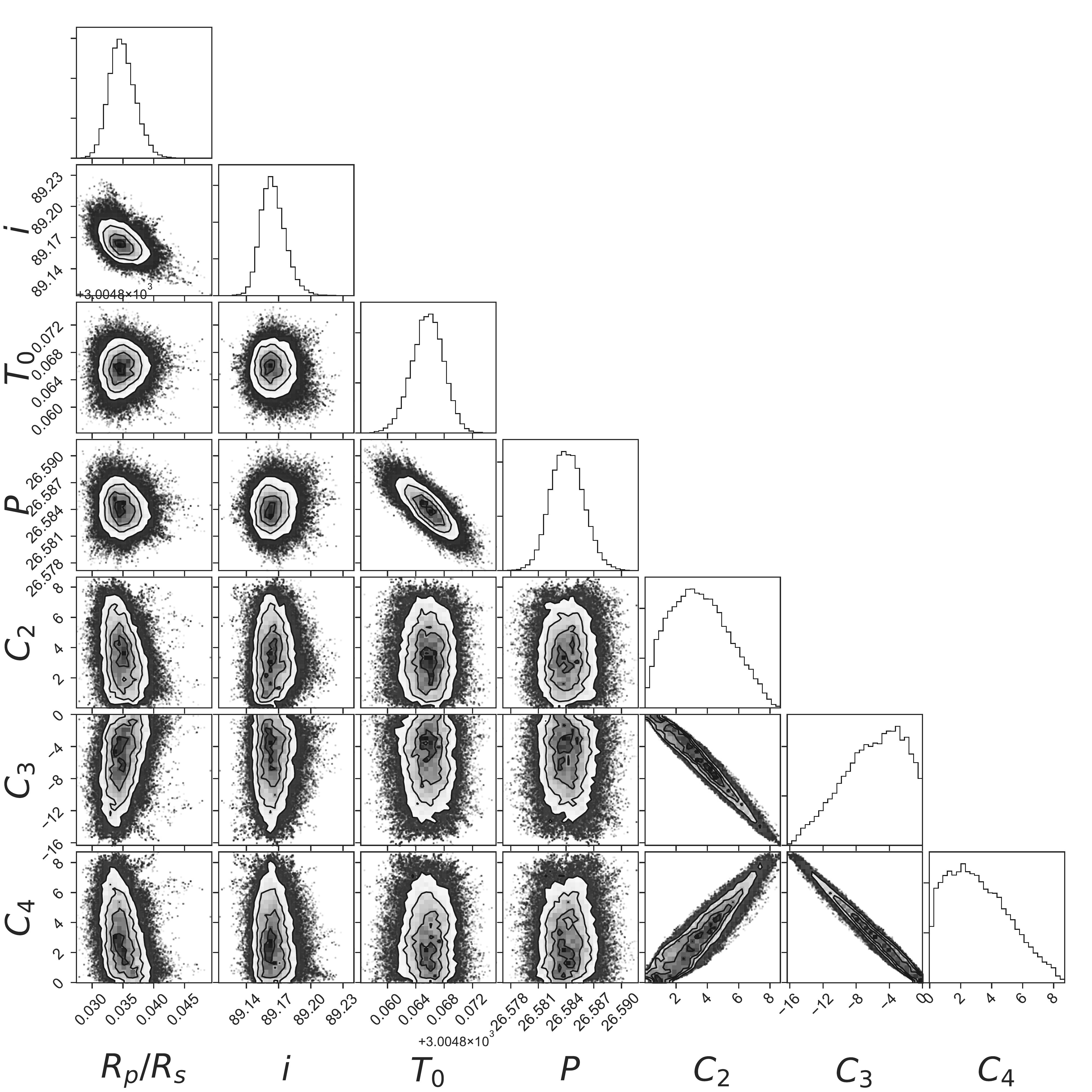}
\caption{MCMC sample distributions from the transit model fit for planet~e. From left to right: planet-star radius ratio, orbital inclination, transit epoch, orbital period and the non-linear limb-darkening coefficients C$_2$, C$_3$ and C$_4$. \label{fig:app-samples}} 
\end{center} 
\end{figure}

We evaluated properties for the newly validated planet by running {\sc batman}/\texttt{emcee} with wide priors on the transit parameters. The phase-folded K2 data and MCMC-fitted transit model are shown in Fig.~\ref{fig:mcmcfit}, and the fitted transit parameters are given in Table~\ref{tab:planete}. All four planet transits were fit simultaneously, allowing the limb-darkening coefficients to vary. The MCMC samples of fitted parameters are shown in Fig.~\ref{fig:app-samples}. Of interest are the planetary radius of 1.7~R$_{\earth}$ and the orbital distance from the host star, which are discussed in terms of potential habitability in Section~\ref{sec:habitability}. We note the lowest flux point has a rather large effect on the fitted parameters -- masking this point leads to a more flat-bottomed transit with a $\chi^2$ value of approximately 1. However, we do not exclude any data points in our analysis.

\begin{table}
\caption{Parameters of planet~e with uncertainties given at the $1 \sigma$ confidence level. $^\ddagger$Computed with supersampling factor = 1. $^\dagger$Assuming an albedo of 0.3.}
\begin{tabular}{lc}
\hline\hline
Property & Value \\
\hline
Fitted properties: & \\ [0.1cm]
T$_{0}$ (BKJD) & $3004.8657_{-0.0023}^{+0.0022}$ \\ [0.1cm]
Period (days) & $26.5841_{-0.0017}^{+0.0018}$ \\ [0.1cm]
R$_{\mathrm{p}}$/R$_{\star}$ & $0.0348_{-0.0019}^{+0.0023}$ \\ [0.1cm]
Inclination (deg.) & $89.164_{-0.010}^{+0.012}$ \\ [0.1cm]
 & \\
Derived properties: & \\ [0.1cm]
Radius (R$_{\earth}$) & $1.73_{-0.13}^{+0.14}$ \\ [0.1cm]
Semi-major axis (R$_{\star}$) & $63.6_{-3.0}^{+3.3}$ \\ [0.1cm]
Semi-major axis (AU) & $0.1346 \pm 0.0011$ \\ [0.1cm]
Impact parameter & $0.928_{-0.045}^{+0.049}$ \\ [0.1cm]
Transit depth (ppm)$^\ddagger$ & $987_{-81}^{+86}$ \\ [0.1cm]
Duration (hours)$^\ddagger$ & $1.44_{-0.06}^{+0.09}$ \\ [0.1cm]
Incident flux (S$_{\earth}$) & $1.80_{-0.23}^{+0.25}$ \\ [0.1cm]
Equilibrium temp. (K)$^\dagger$ & $296 \pm 10$ \\ [0.1cm]
\hline
\end{tabular}
\label{tab:planete}
\end{table}

\section{Discussion}\label{sec:discussion}
\subsection{Temperate Zone}\label{sec:habitability}
Planet~e orbits the host star at a distance of 0.14~AU (64 stellar radii), meaning it receives roughly 1.8 times the stellar flux that Earth does. The $2 \sigma$ confidence of the derived radius means the planet could be equally likely to be rocky or gaseous (1.6 $R_{\earth}$) to very likely to be gaseous (2.0 $R_{\earth}$) \citep{2015ApJ...801...41R}. Therefore the first test to the planets habitability would be to constrain the mass, which will be challenging (see Section~\ref{sec:jwst}). 

Continuing, assuming a rocky planet, we evaluated the stars ``habitable'' zone boundaries using the parametric equations of \citet{2014ApJ...787L..29K}. We find that the planet orbits just interior to the optimistic habitable zone set by the ``recent Venus'' criterion. However, the boundaries predicted by different models vary considerably depending on assumptions and the inclusion of clouds.

\begin{equation}
    \label{eq:esi}
    ESI = 1 - \sqrt{\frac{1}{2} \left(\left(\frac{S - 1}{S + 1}\right)^2 + \left(\frac{R - 1}{R + 1}\right)^2\right)}
\end{equation}

We calculated the Earth Similarity Index\footnote{\url{http://phl.upr.edu/projects/earth-similarity-index-esi}} (ESI; \citealt{2011AsBio..11.1041S}) for planet e using Equation~(\ref{eq:esi}), where $S$ is the incident stellar flux and $R$ is the planetary radius, both relative to Earth. We find a ESI score of 0.72, which places the planet in the top-25 planets most similar to Earth in the Optimistic Sample of the Planetary Habitability Laboratory\footnote{\url{http://phl.upr.edu/projects/habitable-exoplanets-catalog}}.

Due to the tight orbit and likely >1 Gyr age of the system, K2-133~e is expected to be tidally locked to the host star \citep{2009Icar..199..526G} which may reduce the prospect for life. The K2 light curve shows no evidence of stellar flares -- which are capable of stripping atmospheres. The radii of the planets, however, do span the radius gap hinting that their atmospheres may have been eroded in the past and would therefore be uninhabitable. This is discussed more in Section~\ref{sec:evap}.

\subsection{Radius gap}\label{sec:evap}
A radius gap has been identified in the distribution of known exoplanets between 1.5 and 2 $R_{\earth}$ \citep{2017AJ....154..109F,2018AJ....156..264F}. The slope of this gap as a function of orbital period has been found to be negative, which is consistent with the gap caused by photo-evaporation rather than being inherent in planet formation \citep{2018MNRAS.479.4786V}. This would mean planets are formed with gaseous envelopes and then are either stripped to their rocky cores by high stellar X-ray/UV irradiation (shorter periods), or retain their atmospheres in a lower radiation environment (longer periods) while the star becomes less active as it ages \citep{2013ApJ...776....2L,2013ApJ...775..105O}. The core mass of a planet is also important for retaining an atmosphere, as heavier planets will suffer lower mass-loss rates.

The inner three planets in the K2-133 system span the radius gap, with radii of 1.3, 1.6 and 2.0 $R_{\earth}$. The trend of increasing radius with decreasing incident flux hints that the system may have been sculpted by photo-evaporation when the star was young and more active. While the radius of the fourth planet is not well constrained ($2 \sigma$ confidence $1.7 \pm 0.3$ R$_{\earth}$), the radius-period trend of the inner planets indicates that it may have been sufficiently far from the star to keep its primordial atmosphere. However, another explanation could be that the planets have different core masses.

K2-133 is comparable to the GJ-9827 system -- three super-Earths orbiting a late K star with radii 1.6, 1.2 and 2.0 $R_{\earth}$ \citep{2017AJ....154..266N}, which has been studied in great detail with precise radial velocity measurements \citep{2018AJ....155..148T,2018A&A...618A.116P,2019MNRAS.tmp..105R}. A similar follow-up study of K2-133 would be interesting as a comparison to the GJ~9827 system. The induced RV shifts should be similar (${\sim}{2}{+}$ m\,s$^{-1}$), however the star is much fainter (V = 14).

\subsection{Follow-up prospects}\label{sec:jwst}
The system is challenging for follow-up observations, due to the planets producing shallow transits and weak RV shifts. The star is reasonably bright towards the IR (K = 10) but fainter in the optical (V = 14), making ground observations particularly difficult. Ideally follow-up observations would provide information on the planetary masses and atmospheres, plus a stronger constraint on the radius of planet~e or possibly identify additional planets in the system.

It is possible to infer masses from transit timing variations (TTVs) \citep{2018A&A...613A..68G,2016ApJ...820...39J,2013ApJ...777....3N}. This involves precisely measuring the central transit times of individual events which can vary by seconds to hours due to gravitational interactions between the planets. Using the mass-radius relation of \citet{2014ApJ...783L...6W}, we estimate the mid-transit times of the K2-133 planets to vary by 5-40 seconds using the \texttt{TTVFaster} code \citep{2016ApJ...818..177A}, however these amplitudes could be larger due to heavier planets or the presence of more planets in the system. We do not identify any significant TTVs in the K2 data, however due to the cadence we are limited to a $1 \sigma$ transit timing precision of 2.2+ minutes.

K2-133 (TIC~150096001) lies close to the ecliptic plane with an ecliptic latitude of $2.8^{\circ}$. This means it will not be observed in TESS Cycle 2 which will view a minimum latitude of $6^\circ$. However, TESS may still observe the ecliptic in its extended mission. K2-133 has an estimated TESS mag of 12.3 in the TESS Input Catalog \citep{2018AJ....156..102S}, which corresponds to a 1-hour noise level of roughly 750ppm \citep{ticgen}. Individual transits should therefore be detected at ca. 1.5-3 sigma, with three of the planets transiting more than once. New transit data could also be acquired with the forthcoming PLATO mission \citep{2014ExA....38..249R}, which is planned to launch in 2026. K2-133 may be observed in one of the stop-and-stare pointings, which will be defined 2 years before launch \citep{2018haex.bookE..86R}. Unfortunately K2-133 is too faint for observations with CHEOPS, which will have an approximate limiting V-band magnitude of 12 \citep{2013EPJWC..4703005B}.

\section{Conclusions}
We have validated a fourth planet in the K2-133 system. This planet has a radius of $1.73_{-0.13}^{+0.14}$ R$_{\earth}$ and orbits on the inner edge of the stars optimistic habitable zone. We have also presented updated properties of the host star and three already validated planets. The system warrants further study to test theories of the cause of the observed radius gap in the exoplanet population, such as photo-evaporation. Planetary mass measurements are needed for this, which will be a challenge to obtain due to the faintness of the host star.

\section*{Acknowledgements}
We thank the reviewer for helpful comments to improve the manuscript. We also thank Dr. Ian Crossfield and Dr. David Ciardi for allowing the use of their AO imaging observations and analysis. This research made use of the WHT and its service programme, which are operated on the island of La Palma by the Isaac Newton Group of Telescopes in the Spanish Observatorio del Roque de los Muchachos of the Instituto de Astrof\'{\i}sica de Canarias. This research made use of the Exoplanet Follow-up Observation Program website, which is operated by the California Institute of Technology, under contract with the National Aeronautics and Space Administration under the Exoplanet Exploration Program. RW acknowledges funding from the Northern Ireland Department for Education. KP acknowledges funding from the Leibniz Association through its 2019 Leibniz Competition Program. CAW acknowledges support from the STFC grant ST/P000312/1.

\bibliographystyle{mnras}
\bibliography{bib}



\bsp	
\label{lastpage}
\end{document}